\documentstyle[12pt]{article}
\title{{\bf Convolution of Ultradistributions and Field Theory}
\thanks{\it{This work was partially supported by Consejo Nacional de
Investigaciones Cient\'{\i}ficas and Comisi\'{o}n de Investigaciones
Cient\'{\i}ficas de la Pcia. de Buenos Aires; Argentina.}}}
\author{C.G.Bollini${}^1$,T.Escobar${}^2$ and M.C.Rocca${}^1$\\
${}^1$Departamento de F\'{\i}sica, Fac. de Ciencias Exactas,\\
Universidad Nacional de La Plata.\\
C.C. 67 (1900) La Plata. Argentina. \\
${}^2$Departamento de Matem\'atica, Fac. de Ciencias Exactas,\\ 
Universidad Nacional de La Plata.\\
C.C. 172 (1900) La Plata. Argentina.}

\date{May 1, 1998}

\begin{document}

\maketitle

\begin{abstract}

In this work, a general definition of Convolution between two arbitrary
Tempered Ultradistributions is given. When one of the Tempered
Ultradistributions is rapidly decreasing this definition coincides with 
the definition of J. Sebastiao e Silva.

The product of two arbitrary distributions of exponential type is defined 
via the Convolution of its corresponding Fourier Transforms.

Several examples of Convolution of two Tempered Ultradistributions 
and singular products are given. In particular, we reproduce 
the results obtained by A. Gonzales Dominguez and A. Bredimas.


\end{abstract}

\renewcommand{\theequation}{\arabic{section}.\arabic{equation}}

\section{Introduction}

In physics, it is sometimes necessary to work with functions that grow
exponentially in space or time. For those cases Schwartz Space of
Tempered Distributions ( ref.\cite{tp1} ) is too restrictive. On the 
other hand the space of test functions with bounded support allows the
distributions to blow-up more rapidly than any exponential. In this
sense they should be considered to be too ``permissive'' for physical
applications. What is needed is an equilibrium between the necessities
in x-space and the possibility to work in the Fourier transformed 
space ( p-space ) with propagators. The latter are from a mathematical
point of view analytic functionals defined on a space of entire test
functions.

We shall see that a point of equilibrium is achieved by working
with Tempered Ultradistributions ( see below ). They also have the
advantage of being representable by means of analytic functions.
So that, in general, they are easier to work with and have interesting
properties. One of those properties ( as we shall see ) is the
possibility of defining a convolution product which is general
enough to be valid for any two Tempered Ultradistributions, and of 
course, this automatically provided a definition for the product
of Distributions of the Exponential Type in x-space.

In sections 2 and 3 we define the Distributions of Exponential Type
and the Fourier transformed Tempered Ultradistributions. Each of them
is part of a Guelfand's Triplet ( or Rigged Hilbert Space \cite{tp2} )
together with their respectives duals and a ``middle term'' Hilbert
space.

In section 4 we give a general expression for the convolution. We also
state and prove some existence teorems.

In section 5 we present several examples. Some of them imply singular 
products.

Finally, section 6 is reserved for a discussion of the principal results.
For the benefit of the reader an Appendix is added containing some
formulas utilized in the text.

\section{Distributions of Exponential Type}

For the sake of the reader we shall present a brief description of the
principal properties of Tempered Ultradistributions.

The space $H$  of test functions
such that $e^{p|x|}|D^q\phi(x)|$ is bounded for any p and q 
is defined ( ref.\cite{tp3} ) by means 
of the countably set of norms:
\begin{equation}
{\|\hat{\phi}\|}_p^{''}=\sup_{0\leq q\leq p,\,x} 
e^{p|x|} \left|D^q \hat{\phi} (x)\right|\;\;\;,\;\;\;p=0,1,2,...
\end{equation}
According to reference\cite{tp4} $H$ is a ${\cal K}\{M_p\}$ space
with:
\begin{equation}
M_p(x)=e^{(p-1)|x|}\;\;\;,\;\;\; p=1,2,...
\end{equation}
${\cal K}\{e^{(p-1)|x|}\}$ satisfies condition $({\cal N})$
of Guelfand ( ref.\cite{tp5} ). It is a countable Hilbert and nuclear
space:
\begin{equation}
{\cal K}\{e^{(p-1)|x|}\} = H = \bigcap\limits_{p=1}^{\infty} H_p
\end{equation}
where $H_p$ is obtained by completing $H$ with the norm induced by
the scalar product:
\begin{equation}
{<\hat{\phi}, \hat{\psi}>}_p = \int\limits_{-\infty}^{\infty}
e^{2(p-1)|x|} \sum\limits_{q=0}^p D^q \hat{\phi} (x) D^q 
\hat{\psi} (x)\;dx \;\;\;;\;\;\;p=1,2,...
\end{equation}
If we take the usual scalar product:
\begin{equation}
<\hat{\phi}, \hat{\psi}> = \int\limits_{-\infty}^{\infty}
\overline{\hat{\phi}}(x) \hat{\psi}(x)
\end{equation}
then $H$, completed with (2.5), is the Hilbert space ${\cal H}$
of square integrable functions.

The space of continuous linear functionals defined on $H$ is the space
${\Lambda}_{\infty}$ of the distributions of the exponential type 
( ref.\cite{tp3} ).

The ``nested space''
\begin{equation}
\left( H, {\cal H}, {\Lambda}_{\infty} \right)
\end{equation}
is a Guelfand's triplet ( or a Rigged Hilbert space \cite{tp2} ).

Any Guelfand's triplet $( {\cal A}, {\cal H}, {\cal A}^{'} )$
has the fundamental property that a linear and symmetric operator
on ${\cal A}$, admitting an extension to a self-adjoint operator in  
${\cal H}$, has a complete set of generalized eigen-functions 
in ${\cal A}^{'}$ with real eigenvalues.

\section{Tempered Ultradistributions} 

\setcounter{equation}{0}

The Fourier transform of a function $\hat{\phi}\in H$ is
\begin{equation}
\phi(z)=\frac {1} {2\pi}
\int\limits_{-\infty}^{\infty}dx\;e^{izx}\overline{\hat{\phi}}(x)
\end{equation}
$\phi(z)$ is entire analytic and rapidly decreasing on straight lines 
parallel
to the real axis. We shall call $h$ the set of all such functions.
\begin{equation}
h={\cal F}\left\{H\right\}
\end{equation}
It is a ${\cal Z}\{M_p\}$ space ( ref.\cite{tp5} ), countably normed
and complete, with:
\begin{equation}
M_p(z)= (1+|z|)^p
\end{equation}
h is also a nuclear space with norms:
\begin{equation}
{\|\phi\|}_{pn} = \sup_{|Im(z)|\leq n} {\left(1+|z|\right)}^p
|\phi (z)|
\end{equation}
We can define the usual scalar product:
\begin{equation}
<\phi (z), \psi (z)>=\int\limits_{-\infty}^{\infty}
\phi(z) {\psi}_1(z) dz = 
\int\limits_{-\infty}^{\infty} \overline{\hat{\phi}}(x)
\hat{\psi}(x) dx 
\end{equation}
where:
\[{\psi}_1(z)=\int\limits_{-\infty}^{\infty}dx\;
e^{-izx}\hat{\psi}(x)\]
By completing $h$ with the norm induced by (3.5) we get the Hilbert
space of square integrable functions.

The dual of $h$ is the space ${\cal U}$ of tempered ultradistributions
( ref.\cite{tp3} ). In other words, a tempered ultradistribution is
a continuous linear functional defined on the space $h$ of entire
functions rapidly decreasing on stright lines parallel to the real axis.

The set $(h,{\cal H},{\cal U})$ is also a Guelfand's triplet.

${\cal U}$ can also be characterized in the following way 
( ref.\cite{tp3} ): let ${\cal A}$ be the space of all functions 
$F(z)$ such that:

{\Large I}-$F(z)$ is analytic for $\{z\in {\cal C} : |Im(z)|>p\}$.

{\Large II}-$F(z)/z^p$ is bounded continuous  in $\{z\in {\cal C} : 
|Im(z)|\geq p\}$, where $p=0,1,2,...$ depends on $F(z)$.

Let $\Pi$ be the set of all z-dependent polynomials, $z\in {\cal C}$.
Then ${\cal U}$ is the quotient space:

{\Large III}-   ${\cal U}={\cal A}/\Pi$

Due to these properties it is possible to represent any ultradistribution
as ( ref.\cite{tp3} ):
\begin{equation}
F(\phi)=<F(z), \phi(z)>=\oint\limits_{\Gamma} dz\;F(z) \phi(z) 
\end{equation}
where the path $\Gamma$ runs parallel to the real axis from
$-\infty$ to $\infty$ for $Im(z)>\rho$, $\rho>p$ and back from
$\infty$ to $-\infty$ for $Im(z)<-\rho$, $-\rho<-p$.
( $\Gamma$ lies outside a horizontal band of width 2p containing  
all the singularities of $F(z)$ ).

Formula (3.6) will be our fundamental representation for a tempered
ultradistribution. Sometimes use will be made of ``Dirac formula''
for ultradistributions ( ref.\cite{tp6} ):
\begin{equation}
F(z)=\frac {1} {2\pi i}\int\limits_{-\infty}^{\infty} dt\;
\frac {f(t)} {t-z} 
\end{equation}
where the ``density'' $f(t)$ is such that
\begin{equation}
\oint\limits_{\Gamma} dz\; F(z) \phi(z) =
\int\limits_{-\infty}^{\infty} dt\; f(t) \phi(t) 
\end{equation}
While $F(z)$ is analytic on $\Gamma$, the density $f(t)$ is in
general singular, so that the r.h.s. of (3.8) should be interpreted
in the sense of distribution theory.

Another important property of the analytic representation is the fact
that on $\Gamma$, $F(z)$ is bounded by a power of $z$ ( ref.\cite{tp3} ):
\begin{equation}
|F(z)|\leq C|z|^p
\end{equation}
where $C$ and $p$ depend on $F$.

The representation (3.6) makes evident that the addition of a 
polinomial $P(z)$ to $F(z)$ do not alter the ultradistribution:
\[\oint\limits_{\Gamma}dz\;\{F(z)+P(z)\}\phi(z)=
\oint\limits_{\Gamma}dz\;F(z)\phi(z)+\oint\limits_{\Gamma}dz\;
P(z)\phi(z)\]
But:
\[\oint\limits_{\Gamma}dz\;P(z)\phi(z)=0\]
as $P(z)\phi(z)$ is entire analytic ( and rapidly decreasing ),
\begin{equation}
.{}^..\;\;\;\;\oint\limits_{\Gamma}dz\;\{F(z)+P(z)\}\phi(z)=
\oint\limits_{\Gamma}dz\;F(z)\phi(z)
\end{equation}

\section{The Convolution} 

\setcounter{equation}{0}

If we try to define the convolution product by means of the natural
formula:
\begin{equation}
(F\ast G)\{\phi\}= 
\oint\limits_{{\Gamma}_1} 
\oint\limits_{{\Gamma}_2} dk_1\;dk_2\; F(k_1) G(k_2) 
\phi(k_1+k_2)
\end{equation}
we will soon discover that it is not always defined. The reason is simple.
The result of:
\[\oint\limits_{\Gamma}dk\;F(k)\phi(k+k^{'})=\chi(k^{'})\]
does not, in general, belong to $h$. However, if at least one of
the ultradistributions $F$ and $G$ is rapidly decreasing ( say $G$ ),
then a convolution can be defined ( ref.\cite{tp6} ) by:
\begin{equation}
H(k)=\int\limits_{-\infty}^{\infty}dt\;f(t) G(k-t)
\end{equation}
where $f(t)$ is the density associated to $F(k)$ ( cf.(3.7) ).

In order to eliminate the test function from (4.1) use can be made 
of the complex $\delta$-function, which is an ultradistribution
( Cauchy's theorem ):
\begin{equation}
{\delta}_{z^{'}}\{\phi\}=-\frac {1} {2\pi i}
\oint\limits_{\Gamma}dz\; \frac {\phi(z)} {z-z^{'}}=
\phi(z^{'})
\end{equation}
where the point $z^{'}$ is enclosed by $\Gamma$. ( This procedure
was previously used in ref.\cite{tp7} ). We can then write (4.1)
as:
\begin{equation}
(F\ast G)\{\phi\}=-\frac {1} {2\pi i}
\oint\limits_{\Gamma}dz
\oint\limits_{{\Gamma}_1}                                               
\oint\limits_{{\Gamma}_2} dk_1\;dk_2 \;
\frac {F(k_1) G(k_2)} {z-k_1-k_2} \phi(z)
\end{equation}
The path $\Gamma$ must have: 
\begin{equation}
|Im(z)|>|Im(k_1)|+|Im(k_2)|
\end{equation}
in order to embrace the point $k_1+k_2$, ( $k_1\in{\Gamma}_1$,
$k_2\in{\Gamma}_2$ ).

Equation (4.4) leads to:
\begin{equation}
F\ast G=H=-\frac {1} {2\pi i}
\oint\limits_{{\Gamma}_1}                                               
\oint\limits_{{\Gamma}_2} dk_1\;dk_2 \;
\frac {F(k_1) G(k_2)} {z-k_1-k_2}
\end{equation}
However, we do not expect (4.6) to define a tempered ultradistribution
for every pair $F$, $G$. Note that in (4.1) $F$ and $G$ operate on 
$\phi(k)$ which is rapidly decreasing, while in (4.6) they act on
$(z-k)^{-1}$ $(k=k_1+k_2)$. Furthermore, due to (4.5) and the fact
that ${\Gamma}_1$ and ${\Gamma}_2$ run outside a horizontal band
containing all the singularities of $F$ and $G$, the integrand
in (4.6) is analytic at every point of the integration paths.
Taking into account the property (3.9) of tempered ultraditributions,
we come to the conclusion that the integrations in (4.6) have
at most, a tempered singularity for $k\rightarrow\infty$. In order
to control this possible singularity we introduce a regulator
( see ref.\cite{tp8} ).

We define:
\begin{equation}
H_{\lambda}(z)=\frac {i} {2\pi}
\oint\limits_{{\Gamma}_1}                                               
\oint\limits_{{\Gamma}_2} dk_1\;dk_2 \;
\frac {k_1^{\lambda} F(k_1) k_2^{\lambda} G(k_2)}
{z-k_1-k_2}
\end{equation}
Now, if we have the bounds:
\begin{equation}
|F(k_1)|\leq C_1|k_1|^m\;\;\;;\;\;\;|G(k_2)|\leq C_2|k_2|^n
\end{equation}
Then (4.7) is convergent for
\begin{equation}
Re(\lambda)<-l-1\;\;;\;\;l=max\{m,n\}
\end{equation}
It is also analytic in the region (4.9) of the $\lambda$ plane. 
As the derivative
with respect to $\lambda$ merely multiplies by a logarithmic factor the
integrand of (4.7) without spoiling the convergence.

According to the method of ref.\cite{tp8}, $H_{\lambda}$ can be
analytically continued to other parts of the $\lambda$ plane.
In particular near the origin we have the Laurent ( or Taylor )
expansion:
\begin{equation}
H_{\lambda}=\sum\limits_nH^{(n)}(z){\lambda}^n
\end{equation}
where the sum might have terms with negative n. We now define 
the convolution product as the $\lambda$-independent term of (4.10):
\begin{equation}
H(z)=H^{(0)}(z)
\end{equation}
Note that the derivatives of $H_{\lambda}(z)$ with respect to $z$
can be obtained from (4.7) by taking different powers of the 
denominator:
\begin{equation}
\frac {d^pH_{\lambda}(z)} {dz^p}=(-1)^p p!
\frac {i} {2\pi}
\oint\limits_{{\Gamma}_1}                                               
\oint\limits_{{\Gamma}_2} dk_1\;dk_2 \;
\frac {k_1^{\Lambda} F(k_1) k_2^{\Lambda} G(k_2)}
{(z-k_1-k_2)^p}
\end{equation}
The convergence of (4.7) also ensures that of (4.12). Therefore, also
ensure analyticity in $z$, outside the horizontal band defined
by (4.5). We will now show that $|H_{\lambda}(z)|$ is bonded by
a power of $|z|$ ( cf.(3.9) ).

To that aim we take:
\[Im(\lambda)=0\;\;;\;\;\lambda<-l-1\;\;;\;\;z=x+iy\]
\[k_i={\kappa}_i\pm i{\sigma}_i\;\;;\;\;
{\sigma}_i>0\;\;;\;\;dk_i=d{\kappa}_i\]
The integrals along ${\Gamma}_i$, can be expressed as integrals
on $d{\kappa}_i$ between $0\rightarrow\infty$. Then we have:
\[|H_{\lambda}|=\frac {1} {2\pi}\left|
\oint\limits_{{\Gamma}_1} 
\oint\limits_{{\Gamma}_2} dk_1\;dk_2 \;
\frac {k_1^{\lambda} F(k_1) k_2^{\lambda} G(k_2)}
{z-k_1-k_2} \right| \leq \]
\[\frac {1} {2\pi}\oint\limits_{{\Gamma}_1}
\oint\limits_{{\Gamma}_2}sgnIm(k_1)\;dk_1\;sgnIm(k_2)\;dk_2
\frac {|k_1|^{\lambda} C_1 |k_1|^m |k_2|^{\lambda} C_2 |k_2|^n}
{|z-k_1-k_2|}\leq\]
\[\frac {C_1C_2} {2\pi}\oint\limits_{{\Gamma}_1}
\oint\limits_{{\Gamma}_2}sgnIm(k_1)\;dk_1\;sgnIm(k_2)\;dk_2
|k_1|^{{\lambda}+m}|k_2|^{{\lambda}+n}=\]
\begin{equation}
\frac {8C_1C_2} {\pi} 
\int\limits_0^{\infty} \int\limits_0^{\infty} d{\kappa}_1\;d{\kappa}_2
\left({\kappa}_1^2+{\sigma}_1^2\right)^{\frac {\lambda+m} {2}}
\left({\kappa}_2^2+{\sigma}_2^2\right)^{\frac {\lambda+n} {2}}
\end{equation}
We make now the change of variables $w_i={\kappa}_i^2$ and obtain:
\begin{equation}
(4.13)=\frac {2C_1C_2} {\pi} 
\int\limits_0^{\infty}dw_1\;w_1^{-\frac {1} {2}}
\left(w_1+{\sigma}_1^2\right)^{\frac {\lambda+m} {2}}
\int\limits_0^{\infty}dw_2\;w_2^{-\frac {1} {2}}
\left(w_2+{\sigma}_2^2\right)^{\frac {\lambda+m} {2}}=
\end{equation}
\[\frac {2C_1C_2} {\pi}
{\cal B}\left(\frac {1} {2},-\frac {\lambda+m+1} {2}\right)
{\cal B}\left(\frac {1} {2},-\frac {\lambda+n+1} {2}\right)
{\sigma}_1^{\frac {\lambda+m+1} {2}}
{\sigma}_2^{\frac {\lambda+n+1} {2}} \leq\]
\begin{equation}
C(\lambda,m,n)|z|^{\lambda+m+n+1}
\end{equation}
where ${\cal B}(x,y)$ is Gauss beta function.

It is to be noted that if $G(k)$ is a rapidly decreasing ultradistribution,
then $H_{\lambda}(z)$ ( eq.(4.7) ) coincides with $H_0(z)$:
\begin{equation}
H_0(z)=\frac {i} {2\pi}\oint\limits_{{\Gamma}_1}dk_1\;F(k_1)
\oint\limits_{{\Gamma}_2}dk_2\;\frac {G(k_2)} {z-k_1-k_2} 
\end{equation}
In fact, near $\lambda=0$ we have ( $|k|>1$ ):
\begin{equation}
|k^{\lambda}-1|\leq\lambda(2\pi+|ln|k||)|k|^{\lambda}
\end{equation}
\[H_{\lambda}-H_0(z)=\frac {i} {2\pi}\oint\limits_{{\Gamma}_1}
dk_1\;k_1^{\lambda}F(k_1)
\oint\limits_{{\Gamma}_2}dk_2\;(k_2^{\lambda}-1)
\frac {G(k_2)} {z-k_1-k_2}+\] 
\begin{equation}
\frac {i} {2\pi}\oint\limits_{{\Gamma}_1}
dk_1\;(k_1^{\lambda}-1)F(k_1)
\oint\limits_{{\Gamma}_2}dk_2\;
\frac {G(k_2)} {z-k_1-k_2}
\end{equation}
In eq.(4.18) the integrals are convergent, as $G(k)$ and 
$k^{\lambda}G(k)$ are both rapidly decreasing. Furthermore, due to
(4.17) the difference $H_{\lambda}-H_0$ is proportional to
$\lambda$. So that
\begin{equation}
\lim_{\lambda\rightarrow 0}\; [H_{\lambda}-H_0] =0
\end{equation}
Again, when $G(k)$ is rapidly decreasing, the convolution defined
in ref.\cite{tp6}:
\begin{equation}
H(z)=\int\limits_{-\infty}^{\infty}dt\;f(t)G(z-t)
\end{equation}
(where $f(t)$ is given by (3.7),(3.8)), also coincides with (4.16).
To show that (4.16) implies (4.20), we use (3.8) in (4.16)
\[H_0(z)=\frac {i} {2\pi}\int\limits_{-\infty}^{\infty}dt\;f(t)
\oint\limits_{{\Gamma}_2}dk_2\;\frac {G(k_2)} {z-k_1-k_2}\]
But, if $G(t)$ is the density associated to $G(z)$, then
\[\frac {i} {2\pi}\oint\limits_{{\Gamma}_2}dk_2\;
\frac {G(k_2)} {z-t-k_2}=\frac {1} {2\pi i}
\int\limits_{-\infty}^{\infty}dt_2\;\frac {g(t_2)} {t_2-(z-t)}=
G(z-t)\]
I.e.:
\begin{equation}
H_0(z)=H(z)
\end{equation}

\section{Examples}

\setcounter{equation}{0}

In this section we are going to use definition (4.7) to evaluate
convolution of tempered ultradistributions and, indirectly, product
of distributions ( $\in{\Lambda}_{\infty}$ cf.\S 2 ).

The convolution theorem tells that:
\begin{equation}
{\cal F}\left\{f_1(x)f_2(x)\right\}=\frac {1} {2\pi}
{\check{f}}_1(k)\ast{\check{f}}_2(k)
\end{equation}
where
\[\check{f}={\cal F}\left\{f(x)\right\}(k)\]
{\bf {\Large{I)}}}.-As a first example we shall take the distribution
$x_{\pm}^{\alpha}$ ( ref.\cite{tp8}, ch.1, \S 3.2, also ref.\cite{tp9},
ch.4 ) whose Fourier transform we write
\begin{equation}
{\check{x}}_{\pm}^{\alpha}=ie^{\mp i\frac {1} {2} \alpha}
\Gamma(\alpha +1)k^{-\alpha-1}\Theta[\mp\epsilon(k)]
\end{equation}
where $\Theta(x)$ is Heaviside's step function and $\epsilon(k)=
sgnIm(k)$.

The ultradistribution (5.2) has a line of singularities ( a discontinuity )
on the real axis. Then the path $\Gamma$ ( cf. (2.6) ) runs
parallel to the real axis at a distance as small as we please.
\[{\cal F}\left\{x_+^{\alpha}x_+^{\beta}\right\}=
\frac {i} {4{\pi}^2} \oint\limits_{{\Gamma}_1} dk_1
\oint\limits_{{\Gamma}_2}dk_2\;
\frac {{\check{x}}_+^{\alpha}{\check{x}}_+^{\beta}}
{z-k_1-k_2}=\]
\[\left[\frac {i} {4{\pi}^2} i e^{-i\frac {\pi} {2}\alpha}
\Gamma(\alpha+1) i e^{-i\frac {\pi} {2}\beta}
\Gamma(\beta+1)\right]\;\times\]
\[\oint\limits_{{\Gamma}_1}dk_1k_1^{-\alpha-1}\Theta[-\epsilon(k_1)]
\oint\limits_{{\Gamma}_2}dk_2
\frac {k_2^{-\beta-1}\Theta[-\epsilon(k_2)]} {z-k_1-k_2}\]
The functions $\Theta[\epsilon(k_1)]$ and $\Theta[\epsilon(k_2)]$
eliminate the branches of ${\Gamma}_1$ and ${\Gamma}_2$ ( resp. )
on the lower half plane of $k_1$ and $k_2$. By taking the remaining
integration arbitrarily close to the real axis, we get:
\[{\cal F}\left\{x_+^{\alpha}x_+^{\beta}\right\}=- 
[\;]\oint\limits_{{\Gamma}_1}dk_1k_1^{-\alpha-1}\Theta[-\epsilon(k_1)]
\int\limits_{-\infty}^{\infty} dy\;
\frac {(y-i0)^{-\beta-1}} {z-k_1-y}=\]
\[-[\;]\oint\limits_{{\Gamma}_1}dk_1k_1^{-\alpha-1}
\Theta[-\epsilon(k_1)]
\int\limits_{-\infty}^{\infty} dy\;
\frac {y_+^{-\beta-1}+e^{-i\pi(-\beta-1)}
y_-^{-\beta-1}}{z-k_1-y}=\]
\[-[\;]\oint\limits_{{\Gamma}_1}dk_1k_1^{-\alpha-1}
\Theta[-\epsilon(k_1)]
\frac {\Gamma(-\beta)\Gamma(1+\beta)}
{(z-k_1)^{\beta+1}}\;\times \]
\[\left[e^{-i\pi(-\beta-1)}-
e^{-i\pi\epsilon(z)(-\beta-1)}\right]=\]
\[2i[\;]\Theta[-\epsilon(z)]\Gamma(-\beta)
\Gamma(1+\beta)\sin\pi(-\beta-1)\;\times\]
\[\int\limits_{{\Gamma}_1}dk_1\;\frac {k_1^{-\alpha-1}}
{(z-k_1)^{\beta+1}}\Theta[-\epsilon(k_1)]=\]
\[2i\pi\Theta[-\epsilon(z)][\;]\int\limits_{-\infty}^{\infty}dx\;
\frac {x_+^{-\alpha-1}+
e^{-i\pi(-\alpha-1)}x_-^{-\alpha-1}}
{(z-x)^{\beta+1}}=\]
\[2i\pi\Theta[-\epsilon(z)][\;]
{\cal B}(-\alpha,\beta+\alpha+1)
\left[e^{i\pi\epsilon(z)\alpha
}-e^{i\pi\alpha}\right]z^{-\alpha-\beta-1}=\]
\[2i\pi\{\Theta[-\epsilon(z)]\}^2[\;]
\frac {\Gamma(-\alpha)
\Gamma(\beta+\alpha+1)} {\Gamma(\beta+1)}
2i\sin\pi(-\alpha) z^{-\alpha-\beta-1}=\]
\[ie^{-i\frac {\pi} {2}(\alpha+\beta)}\Gamma(\alpha+\beta+1)
z^{-\alpha-\beta-1}\Theta[-\epsilon(z)]=\] 
\begin{equation}
{\check{x}}_+^{\alpha+\beta}={\cal F}\left\{x_+^{\alpha+\beta}\right\}
={\cal F}\left\{x_+^{\alpha}x_+^{\beta}\right\}
\end{equation}
Where use has been made of equation (A.4) of the
Appendix.

For the evaluation of the convolution ${\check{x}}_+^{\alpha}\ast
{\check{x}}_-^{\beta}$ the procedure is entirely similar. However,
in this case one of the integrations gives rise to a factor 
$\Theta[-\epsilon(z)]$ and the other to a factor $\Theta[\epsilon(z)]$.
So that, instead of $\{\Theta[-\epsilon(z)]\}^2=\Theta[-\epsilon(z)]$, 
we get $\Theta[-\epsilon(z)]\Theta[\epsilon(z)]=0$. I.e.
\begin{equation}
{\check{x}}_+^{\alpha}\ast {\check{x}}_-^{\beta}\equiv 0 
\;\;\;.{}^. .\;\;\; x_+^{\alpha}\cdot x_-^{\beta}=0
\end{equation}
{\bf {\Large{II)}}}.-As a second example we consider Dirac's
$\delta$-functions, whose Fourier transform is:
\begin{equation}
{\check{\delta}}^{(m)}=i^mk^m\frac {\epsilon(k)} {2}
\end{equation}
For the convolution (4.7) we have:
\[{\check{\delta}}^{(m)}\ast{\check{\delta}}^{(n)}=\frac {i} {4\pi}
\int\limits_{{\Gamma}_1}dk_1\;i^mk_1^{\lambda+m}
\frac {\epsilon(k_1)} {2}
\int\limits_{{\Gamma}_2}dk_2\;
\frac {i^nk_2^{\lambda+n}\epsilon(k_2)} {z-k_1-k_2}=\]
In this case, the factors ${\epsilon}_1$ and ${\epsilon}_2$
change the sign of integrations of the lower half plane of $k_1$
and $k_2$.
\[\frac {i^{m+n+1}} {4\pi}
\int\limits_{{\Gamma}_1}dk_1\;k_1^{\lambda+m}
\frac {\epsilon(k_1)} {2}
\int\limits_{-\infty}^{\infty}dy\;
\frac {(y+i0)^{\lambda+n}+(y-i0)^{\lambda+n}} {z-k_1-y}=\]
\[\frac {i^{m+n+1}} {2\pi}
\int\limits_{{\Gamma}_1}dk_1\;k_1^{\lambda+m}
\frac {\epsilon(k_1)} {2}
\int\limits_{-\infty}^{\infty}dy\;
\frac {y_+^{\lambda+n}+\cos\pi(\lambda+n) y_-^{\lambda+n}} 
{z-k_1-y}=\]
\[\frac {i^{m+n+1}} {2\pi}
\int\limits_{{\Gamma}_1}dk_1\;k_1^{\lambda+m}
\frac {\epsilon(k_1)} {2}
\frac {\Gamma(\lambda+n+1)\Gamma(-\lambda-n)} 
{z-k_1}
\left[\cos\pi(\lambda+n)-e^{-i\pi\epsilon(z)(\lambda+n)}\right]=\]
\[-\frac {i\pi\epsilon(z)} {2\pi} i^{m+n+1}
\int\limits_{-\infty}^{\infty}dx\;
\frac {x_+^{\lambda+m}+\cos\pi(\lambda+m)x_-^{\lambda+m}}
{(z-x)^{-\lambda-n}}=\]
\[\frac {\epsilon(z)} {2} i^{m+n}
\frac {\Gamma(\lambda+m+1)\Gamma(-2\lambda-m-n-1)}
{\Gamma(-\lambda-n)}z^{2\lambda+m+n+1}\;\times\]
\[\left[e^{-i\pi\epsilon(z)(\lambda+m+1})+\cos\pi(\lambda+m)\right]=\]
\[\frac {[\epsilon(z)]^2} {2} i^{m+n+1}
\frac {\Gamma(\lambda+m+1)\Gamma(-2\lambda-m-n-1)}
{\Gamma(-\lambda-n)}\sin\pi(\lambda+m)
z^{2\lambda+m+n+1}=\]
\begin{equation}
{}_{\lambda\rightarrow 0}\longrightarrow 0=
{\check{\delta}}^{(m)}\ast{\check{\delta}}^{(n)}
\end{equation}
There are two reasons for this null result. The $\Gamma$ functions 
have simple poles when their arguments are negative integer ( or zero ).
So that the quotient of $\Gamma$ functions has a finite limit.
However they are multiplied by 
$\sin\pi(\lambda+m){}_{\lambda\rightarrow 0}\longrightarrow 0$.

Furthermore, $[\epsilon(z)]^2=1$, and
\[z^{2\lambda+m+n+1}\;\;\;
{}_{\lambda\rightarrow 0}\longrightarrow \;\;\;
z^{m+n+1}\]
Then we can put ( $C$ = arbitrary constant )
\begin{equation}
{\check{\delta}}^{(m)}\ast{\check{\delta}}^{(n)}= C z^{m+n+1}
\end{equation}
But due to the property III, $\S 3$, the ultradistribution (5.7)
is equivalent to zero.

We have then:
\begin{equation}
{\delta}^{(m)}(x)\cdot{\delta}^{(n)}(x)=0
\end{equation}
This result was previously obtained in ref.\cite{tp10} and can be
summarized in general as:

``The product of two distributions with point support is zero''.

{\bf {\Large{III)}}}.-We can combine examples I and II, to find 
the product ${\delta}^{(m)}\cdot{\check{x}}_+^{\alpha}$.
\[\frac {1} {2\pi}{\check{\delta}}^{(m)}\ast
{\check{x}}_+^{\alpha}=\left[\frac {i} {4{\pi}^2}i^mi 
e^{-i\frac {\pi} {2}\alpha}\Gamma(\alpha+1)\right]\;\times\]
\[\oint\limits_{{\Gamma}_1}dk_1\;k_1^{\lambda+m}
\frac {\epsilon(k_1)} {2} 
\oint\limits_{{\Gamma}_2}dk_2\;\frac {k_2^{-\alpha-1}
\Theta[-\epsilon(k_2)]} {z-k_1-k_2}=\]
\[2\pi i\Theta[-\epsilon(z)][\;]\int\limits_{-\infty}^{\infty}dx\;
\frac {x_+^{\lambda+m}+\cos\pi(\lambda+m)x_-^{\lambda+m}}
{(z-x)^{\alpha+1}}=\]
\[2\pi i\Theta[-\epsilon(z)][\;] 
\frac {\Gamma(\lambda+m+1)\Gamma(\alpha-\lambda-m)}
{\Gamma(\alpha+1)}z^{\lambda+m-\alpha}\;\times\]
\[\left[e^{-i\pi\epsilon(z)(\lambda+m+1)}+\cos\pi(\lambda+m)\right]=\]
\[2\pi i\Theta[-\epsilon(z)] \left(-\frac {i^m} {4{\pi}^2}
e^{-i\frac {\pi} {2}\alpha}\right)\Gamma(\lambda+m+1)
\Gamma(\alpha-\lambda-m)\;\times\]
\begin{equation}
i\sin\pi(\lambda+m)\;\epsilon(z)z^{\lambda+m-\alpha}
{\longrightarrow 0}_{\lambda\rightarrow 0}\;,
\end{equation}
if $\alpha$ is not an integer $s$ such that $s\leq m$.

When $0\leq \alpha=s\leq m$:
\[\frac {1} {2\pi} {\check{\delta}}^{(m)}\ast{\check{x}}_+^s=
-2i\pi\Theta[-\epsilon(z)]\frac {i^m} {4{\pi}^2}(-i)^s
i\epsilon(z) z^{\lambda+m-s}\;\times\]
\[\frac {\Gamma(\lambda+m+1)} {\Gamma(\lambda+m+1-s)}
\frac {\sin\pi(\lambda+m)} {\sin\pi(\lambda+m-s)}=\]
\[\frac {i^m} {2} (-i)^s
\frac {\Gamma(\lambda+m+1)} {\Gamma(\lambda+m+1-s)}
\frac {\sin\pi(\lambda+m)} {\sin\pi(\lambda+m-s)}
\Theta[-\epsilon(z)]\epsilon(z)z^{\lambda+m-s}\]
\[{}_{\lambda\rightarrow 0}\longrightarrow\;\;
(-1)^s\frac {i^{m-s}} {2} \frac {m!} {(m-s)!}
\frac {\epsilon(z)} {2} z^{m-s}=\]
\begin{equation}
\frac {(-1)^s} {2} \frac {m!} {(m-s)!}{\check{\delta}}^{(m-s)}
\end{equation}
In particular, for $s=0$ we get:
\begin{equation}
{\delta}^{(m)}(x) \Theta(x)=\frac {1} {2} {\delta}^{(m)}(x)
\end{equation}
If $\alpha=s$=negative number=$-n$ we must be careful as $x_+^{\alpha}$
has a pole for $\alpha=-n$. We shall deal with this case by the 
replacement $\alpha=-n-\lambda$ in (5.9).
\[\Gamma(\alpha-\lambda-m)\longrightarrow\Gamma(-2\lambda-m-n)=\]
\[-\frac {\pi} {\Gamma(2\lambda+m+n+1)\sin\pi(2\lambda+m+n)}\]
And by taking the limit $\lambda\rightarrow 0$:
\[\frac {1} {2\pi} {\check{\delta}}^{(m)}\ast x_+^{-n}=
\frac {i^{m+n}} {2} \frac {m!} {(m+n)!}\frac {(-1)^n} {2}
\frac {\epsilon(z)} {2}z^{m+n}=\]
\begin{equation}
\frac {(-1)^n} {4} \frac {m!} {(m+n)!}{\check{\delta}}^{(m+n)}
\end{equation}
In eqs. (5.10) and (5.12) we have used:
\[\Theta[-\epsilon(z)]\epsilon(z)=-\Theta[-\epsilon(z)]=
\frac {1} {2} (\epsilon(z)-1)=\frac {\epsilon(z)} {2}-
\frac {1} {2} \]
\[\Theta[-\epsilon(z)]\epsilon(z)z^s=\frac {\epsilon(z)} {2}
z^s-\frac {1} {2} z^s\approx \frac {\epsilon(z)} {2} z^s\]
As $Cz^s$ is equivalent to zero ( cf. (5.7) ).

There are also similar expresions which originates in the use of
${\check{x}}_-^{\alpha}$ in (5.9). In particular, if we use
\begin{equation}
{\check{x}}^{-n}={\check{x}}_+^{-n}+(-1)^n{\check{x}}_-^{-n}
\end{equation}
then we easily find
\begin{equation}
\frac {1} {2\pi} {\check{\delta}}^{(m)}\ast {\check{x}}^{-n}=
\frac {(-1)^n} {2} \frac {m!} {(m+n)!} {\check{\delta}}^{(m+n)}
\end{equation}
The case $m=0$, $n=1$, was first published in ref.\cite{tp11}.
For $m=n$ eq.(5.14) conicides with ref.\cite{tp12}.

{\bf {\Large{IV)}}}.-To illustrate the use of (4.10) and (4.11), we are
now going to examine an interesting example.

Let us take the ultradistribution (5.13), which is found to be:
\begin{equation}
{\check{x}}^{-n}=\frac {(-i)^n\pi} {(n-1)!} \left[-\frac {1} {\pi i} 
\ln(k)+\frac {\epsilon(k)} {2}\right]k^{n-1}
\end{equation}
The convolution product is now:
\[{\check{x}}^{-m}\ast{\check{x}}^{-n}=-\frac {(-i)^{m+n+1}} 
{4(m-1)!(n-1)!}
\oint\limits_{{\Gamma}_1}\oint\limits_{{\Gamma}_2}dk_1\;dk_2\]
\[\left\{-\frac {1} {{\pi}^2}
\frac {k_1^{\lambda+m-1}\ln(k_1)k_2^{\lambda+n-1}\ln(k_2)}
{z-k_1-k_2}
-\frac {1} {2\pi 1}
\frac {k_1^{\lambda+m-1}\ln(k_1)k_2^{\lambda+n-1}\epsilon(k_2)}
{z-k_1-k_2}\;+\right.\]
\begin{equation}
\left.-\frac {1} {2\pi i}
\frac {k_1^{\lambda+m-1}\epsilon(k_1)k_2^{\lambda+n-1}\ln(k_2)}
{z-k_1-k_2}+
\frac {1} {4}
\frac {k_1^{\lambda+m-1}\epsilon(k_1)k_2^{\lambda+n-1}\epsilon(k_2)}
{z-k_1-k_2}\right\}
\end{equation}
The last term of (5.16) is null according to example II). We analyze
now the first term. We shall use the identity
\[k^{\lambda+m-1}\ln(k)=D_{\alpha}k^{\alpha+m-1}\;\;\;;\;\;\;
D_{\alpha}=\left.\frac {\partial} {\partial\alpha}  
\right|_{\alpha=\lambda}\]
Then we have:
\[\frac {i} {4{\pi}^2}\frac {(-i)^{m+n}} 
{(m-1)!(n-1)!}
\oint\limits_{{\Gamma}_1}\oint\limits_{{\Gamma}_2}dk_1\;dk_2\;
\frac {k_1^{\lambda+m-1}\ln(k_1)k_2^{\lambda+n-1}\ln(k_2)}
{z-k_1-k_2}=\]
\[\left[\frac {i} {4{\pi}^2}\frac {(-i)^{m+n}} 
{(m-1)!(n-1)!}\right]
\oint\limits_{{\Gamma}_1}dk_1\;
D_{\alpha}k^{\alpha+m-1}
\oint\limits_{{\Gamma}_2}dk_2\;
\frac {D_{\beta}k_2^{\beta+n-1}\ln(k_2)}
{z-k_1-k_2}=\]
\[[\;]D_{\alpha}D_{\beta}
\oint\limits_{{\Gamma}_1}dk_1\;
\frac {k_1^{\alpha+m-1}} {(z-k_1)^{1-\beta-n}}
2i\sin\pi(\beta+n-1)\Gamma(\beta+n)\Gamma(1-\beta-n)=\]
\[2\pi i[\;]D_{\alpha}D_{\beta}
\oint\limits_{{\Gamma}_1}dk_1\;
\frac {k_1^{\alpha+m-1}} {(z-k_1)^{1-\beta-n}}= \]
\[2\pi i[\;]D_{\alpha}D_{\beta}
\frac {\Gamma(\alpha+m)\Gamma(1-\alpha-m-\beta-n)}
{\Gamma(1-\beta-n}2i\sin\pi(\alpha+m-1)\;z^{\alpha+\beta+m+n-1}=\]
\[4\pi[\;]D_{\alpha}D_{\beta}
\frac {\Gamma(\alpha+m)\Gamma(\beta+n)}
{\Gamma(\alpha+\beta+n+m}
\frac {\sin\pi\alpha\;\sin\pi\beta} {\sin\pi(\alpha+\beta)}
z^{\alpha+\beta+m+n-1}=\]
\begin{equation}
-\frac {1} {\pi} \frac {(-i)^{m+n-1}} {(m+n-1)!}D_{\alpha}D_{\beta}
\left\{
\frac {\sin\pi\alpha\;\sin\pi\beta} {\sin\pi(\alpha+\beta)}
z^{\alpha+\beta+m+n-1}\right\}
\end{equation}
where we have used the fact that any derivative, $D_{\alpha}$ or
$D_{\beta}$ acting on a $\Gamma$ function will lead to a null result
of (5.17) through the substitutions $\alpha=\lambda$, $\beta=\lambda$,
$\lambda\rightarrow 0$. Now the derivatives in (5.17) give raise
essentially to two types of terms. The two derivatives acting on the
trigonometric functions give raise to a pole term ( in $\lambda$ ).
If one takes a derivative of the trigonometric functions and a 
derivative of $z^{\alpha+\beta}$, a constant term is obtained.
For the term $D_{\alpha}D_{\beta}z^{\alpha+\beta}$, the limit
$\lambda\rightarrow 0$ of the trigonometric functions is zero.
Thus we get:
\[(5.17)=-\frac {(-i)^{m+n-1}} {(m+n-1)!} z^{m+n-1}
\left\{\frac {1} {4}\frac {1} {\lambda} z^{2\lambda} +\frac {1} {2}
\ln(z)\right\}\]
The second and third terms of (5.16) have the same contribution,
and can be evaluated by a similar procedure. 
This contribution is:
\[\frac {1} {8\pi}\frac {(-i)^{m+n-2}} {(m-1)!(n-1)!}
\oint\limits_{{\Gamma}_1}\oint\limits_{{\Gamma}_2}dk_1\;dk_2
\frac {k_1^{\lambda+m-1}\ln(k_1)k_2^{\lambda+n-1}\epsilon(k_2)}
{z-k_1-k_2}=\]
\begin{equation}
\frac {(-i)^{m+n}} {(m+n-1)!}\frac {\pi} {4}
\epsilon(z)\;z^{m+n-1}
\end{equation}
According to (5.17) and (5.18), we finally get:
\[(5.16)=\frac {(-i)^{m+n}} {(m+n-1)!} z^{m+n-1}
\left\{\frac {i} {4}\frac {1} {\lambda} z^{2\lambda} +\frac {i} {2}
\ln(z)+ \frac {\pi} {2} \epsilon(z)\right\}=\]
\[\frac {(-i)^{m+n}} {(m+n-1)!} z^{m+n-1}
\left\{\frac {i} {4}\frac {1} {\lambda} 
(1+2\lambda\ln(z)) +\frac {i} {2}
\ln(z)+ \frac {\pi} {2} \epsilon(z)\right\}=\]
\begin{equation}
\frac {(-i)^{m+n}\pi} {(m+n-1)!} z^{m+n-1}
\left\{-\frac {1} {\pi i}
\ln(z)+ \frac {1} {2} \epsilon(z)\right\}
\end{equation}
The $\lambda$-independent term is recognized to be ${\check{x}}^{-m-n}$
(cf.5.15). The pole term is equivalent to zero, according to $\S 3$ III.

{\bf {\Large{V)}}}.-Finally, we give a physical example. We consider
a massless scalar $\frac {\lambda} {4!} {\phi}^4(x)$ theory in four 
dimensions. For this
theory we shall evaluate the self-energy Green function. 

The propagator for the field $\phi(x)$ is ( ref.\cite{tp9}:
\begin{equation}
\Delta(x)= [-4{\pi}^2 (u^2-i0)]^{-1} 
\end{equation}
According to eq.(A.5)-(A.10) of Appendix we can write:
\[(u^2-i0)^{-1} = (2k_0)^{-1}\left[(k_0-r)^{-1}+(k_0+r)^{-1}\right]+\]
\begin{equation}
(2r)^{-1}\left[{\delta}(k_0-r)+{\delta}(k_0+r)\right]+
C{\delta}(k_0-r){\delta}(k_0+r)
\end{equation}
( where $C$ is an arbitrary constant appearing in the definition of some
distributions, ref.\cite{tp9}, 8.8, 8.9
( See also Appendix )).

And using the results of I) to IV) it is easy show that:
\[(u^2-i0)^{-1}(u^2-i0)^{-1}=(u^2-i0)^{-2}\]
Then, we have for the self-energy
\begin{equation}
\Sigma(x)= \left(\Delta(x)\right)^2 =
\frac {1} {16{\pi}^4}(u^2-i0)^{-2}
\end{equation}

Where $(u^2-i0)^{-2}$ is defined in ref.\cite{tp9}, 8.8, 8.9.

\section{Discussion}

When we use the perturbative development in Quantum Field Theory, we
have to deal with products of distributions in configuration space,
or else, with convolutions in the Fourier transformed p-space.
Unfortunately, products or convolutions ( of distirbutions ) are
in general ill-defined quantities. However, in physical applications
one introduces some ``regularization'' scheme, which allows us to
give sense to divergent integrals. Among these procedures we would
like to mention the dimensional regularization method ( ref. 
\cite{tp14,tp15} ). Essentially, the method consist in the
separation of the volume element ( $d^{\nu}p$ ) into an angular
factor ( $d\Omega$ ) and a radial factor ( $p^{\nu-1}dp$ ). 
First the angular integration is carried out and then the number
of dimensions $\nu$ is taken as a free parameter. It can be adjusted
to give a convergent integral, which is an analytic function of
$\nu$. 

Our formula (4.7) is similar to the expression one obtains with
dimensional regularization. However, the parameter $\lambda$ is 
completely independent of any dimensional interpretation.

All ultradistributions provide integrands ( in (4.7) ) that are
analytic functions along the integration path. The parameter
$\lambda$ permit us to control the possible tempered asymptotic
behavior ( cf. eq. (3.9) ). The existence of a region of
analiticity for $\lambda$, and a subsequent continuation to 
the point of interest ( ref. \cite{tp8} ), defines the convolution
product.

Those properties show that tempered ultradistributions provide an
appropriate framework for applications to physics. Furthermore,
they can ``absorb'' arbitrary polynomials, thanks to eq. (3.10). 
A property that is interesting for renormalization theory. ( See
for example the elimination of the pole term in (5.19) ). Consequently,
we began this paper with a summary of the main characteristics 
of tempered ultardistributions and their Fourier transformed 
distributions of the exponential type. 

\setcounter{equation}{0}

\newpage

\section*{Appendix}

\subsection*{Definitions}

\setcounter{section}{1}

\setcounter{subsection}{0}

\setcounter{equation}{0}

\renewcommand{\theequation}{\Alph{section}.
\arabic{equation}}
From ref.\cite{tp8} we quote the formula:
\[{\cal B}(\lambda,\mu)=\int\limits_0^{1/2}dx\;x^{\lambda-1}
\left[(1-x)^{\mu-1}-\sum\limits_{r=0}^{k-1}(-1)^r
\frac {\Gamma(\mu)x^r} {r!\Gamma(\mu-r)}\right]+\]
\[\int\limits_{1/2}^1dx\;(1-x)^{\mu-1}
\left[x^{\lambda-1}-\sum\limits_{r=0}^{s-1}(-1)^r
\frac {\Gamma(\lambda)(1-x)^r} {r!\Gamma(\lambda-r)}\right]+\]
\begin{equation}
\sum\limits_{r=0}^{k-1}\frac {(-1)^r\Gamma(\mu)}
{2^{r+\lambda}r!\Gamma(\mu-r)(r+\lambda)}+
\sum\limits_{r=0}^{s-1}\frac {(-1)^r\Gamma(\lambda)}
{2^{r+\mu}r!\Gamma(\lambda-r)(r+\mu)}
\end{equation}
valid for $Re\;\lambda>-k$, $Re\;\mu>-s$, where $k$ and $s$
are positive integers.\\
From ref.\cite{tp13} we get:
\begin{equation}
{\cal B}(\lambda,\mu)=\frac {\Gamma(\lambda)\Gamma(\mu)}
{\Gamma(\lambda+\mu)}
\end{equation}
\begin{equation}
\Gamma(\lambda)=\int\limits_0^{\infty} dt\;t^{\lambda-1} e^{-t}
\end{equation}
\begin{equation}
\Gamma(\lambda)\Gamma(1-\lambda)=\frac {\pi} {\sin\pi\lambda}
\end{equation}
From ref.\cite{tp9} we have:
\[{\delta}^{(m)}(u^2)={\delta}^{(m)}(x^0+r) (x^0-r)^{-m-1}
sgn(x^0-r)+\]
\begin{equation}
{\delta}^{(m)}(x^0-r) (x^0+r)^{-m-1}
sgn(x^0+r)
\end{equation}
where:
\begin{equation}
u^2=x_0^2-x_1^2- \cdot\cdot\cdot-x_{n-1}^2
\end{equation}
\begin{equation}
r^2=x_1^2+x_2^2+\cdot\cdot\cdot+x_{n-1}^2
\end{equation}
\begin{equation}
(u^2\pm i0)^{-m}=u^{-2m}\pm \frac {(-1)^m} {(m-1)!}
i\pi{\delta}^{(m-1)}(u^2)
\end{equation}
\begin{equation}
x^{-m}sgn(x)=\frac {(-1)^{m-1}} {(m-1)!}\{|x|^{-1}\}^{(m-1)}
\end{equation}
\begin{equation}
|x|^{-1}=\{sgn(x)\ln|x|\}^{'}+C\delta(x)
\end{equation}
where $C$ is an arbitrary constant.

\newpage

\end{document}